\documentclass[journal]{IEEEtran}\usepackage{graphicx}

\usepackage{amsmath}
\usepackage{graphicx}
\usepackage{tikz}
\usepackage{pgfplots}
\usepackage{mathtools}
\usepgfplotslibrary{colorbrewer}
\usepgfplotslibrary{groupplots}
\usetikzlibrary{shapes.geometric}
\usepackage{pgfplotstable}
\pgfplotsset{compat=1.17}
\usetikzlibrary{backgrounds}
\usepackage{wrapfig}  
\usepackage{subfigure}
\usepackage{amsfonts}

\DeclareMathOperator{\asinh}{asinh} 
\DeclareMathOperator{\atan}{atan}
\DeclareMathOperator{\sign}{sign}

\begin{document}

\title{A Closed-form Expression for the ISRS GN Model Supporting Distributed Raman Amplification}

\author{H.~Buglia, M. Jarmolovicius, A. Vasylchenkova, E. Sillekens, R.I. Killey,
        P.~Bayvel,
        and~L.~Galdino
\thanks{This work is partly funded by the EPSRC Programme Grant TRANSNET (EP/R035342/1).  H.~Buglia is funded jointly by an EPSRC studentship EP/T517793/1 and the Microsoft 'Optics for the Cloud' Alliance. A.~Vasylchenkova acknowledges the support of the Leverhulme Trust Early Career Fellowship (ECF-2020-150).}
\thanks{The authors are with the Optical Networks Group, University College London, Department of Electronic and Electrical Engineering, Roberts Building, Torrington Place, London WC1E 7JE, UK
WC1E 7JE, U.K. L. Galdino is with Corning Optical Communications, Ewloe, U.K (e-mail: \{henrique.buglia.20; min.jarmolovicius.17; a.vasylchenkova; e.sillekens; r.killey; p.bayvel;\}@ucl.ac.uk, galdinol@corning.com)}
}

\maketitle

\begin{abstract}
A closed-form model for the nonlinear interference in distributed Raman amplified links is presented, the formula accounts for both forward and backward pumping.
\end{abstract}

\IEEEpeerreviewmaketitle

\section{Introduction}

\IEEEPARstart{U}{ltra-wideband} (UWB) transmission has attracted considerable attention in recent years as a cost-effective solution to satisfy the ever-increasing volumes of data traffic. To achieve real-time prediction of the performance of UWB optical fibre transmission systems, approximations in closed form are used. Of particular interest there are closed-form expressions derived using the inter-channel stimulated Raman scattering (ISRS) Gaussian noise (GN) model~\cite{isrsgnmodel}, due to their simplicity and efficiency in estimating the NLI in UWB systems. This model includes the effect of ISRS and approximations in closed-form have been developed in~\cite{closed_gauss_daniel,ISRSGNmodel_correction,generalizedclosed}. However, among the weaknesses of these models are that the closed-form expressions do not support distributed Raman amplification (DRA), as they were developed for lumped amplifier solutions. A closed-form expression supporting forward (FW) Raman amplification in the presence of ISRS has recently been proposed in~\cite{MZraman}, however, it was validated only over C-band systems.

In this work, we have developed a closed-form expression of the ISRS GN model~\cite{isrsgnmodel} supporting both FW-DRA and backward (BW) DRA. This was enabled by deriving for the first time a semi-analytical solution to model the signal profile in the presence of DRA and ISRS. The proposed closed-form formulation is valid for Gaussian constellations and supports an arbitrary number of Raman pumps. This work also represents the first closed-form expression supporting both FW-DRA and BW-DRA in the presence of ISRS.

\section{The closed-form expression}
The signal-to-noise ratio for the $i$th channel ($\text{SNR}_{i}$) at the end of the span after amplification can be estimated as $\text{SNR}_{i}^{-1} \approx \text{SNR}_{\text{NLI},i}^{-1} + \text{SNR}_{\text{ASE},i}^{-1}+ \text{SNR}_{\text{TRX},i}^{-1}$, where 
$\text{SNR}_{\text{NLI},i}$, $\text{SNR}_{\text{ASE},i}$ $\text{SNR}_{\text{TRX},i}$ originate from fibre nonlinearity, amplifier noise and transceiver noise, respectively. This work is devoted to the calculation of $\text{SNR}_{\text{NLI},i}$. The first step in the derivation of the closed-form expression is to find a suitable function to represent the signal power evolution along the fibre distance in the presence of DRA. To that end, a formula similar to the one proposed in~\cite{zirngibl1998analytical} is derived accounting for FW and BW Raman pumps; this formula is then approximated using a first order Taylor expansion (see Sec.\ref{derivation}), such that the normalised signal profile $\rho(f_i,z) = P(z,f_i)/P(L,f_i)$ is represented as $\rho(z,f_i) = e^{-\alpha z} [1 - (C_f P_{f}L_{eff} + C_b P_{b} \tilde{L}_{eff})(f_i - \hat{f})]$, with $L_{eff}(\zeta) = (1-e^{-\alpha_f z})/\alpha_f$ and  $\tilde{L}_{eff}(\zeta) = (e^{-\alpha_b(L-z)}-e^{-\alpha_b L})/\alpha_b$, where $f_i$ is the frequency of the channel of interest, $L$ is the span length, $\alpha$, $\alpha_f$ and $\alpha_b$ are the fibre attenuation at the signal, FW- and BW-DRA wavelengths, respectively, $\hat{f}$ is the average frequency of the FW and BW pumps, $P_{f}$, and $P_{b}$ are the total launch power respectively from the WDM channels together with any FW pumps, and the BW pumps, $C_f$ and $C_b$ is the slope of a linear regression of the normalized Raman gain spectrum. 
The coefficients $\alpha$, $C_f$, $C_b$, $\alpha_f$, $\alpha_b$ are channel-dependent parameters and matched using nonlinear least-squares fitting to reproduce the solution of the Raman differential equations in the presence of DRA. These parameters model respectively the fibre loss, the gain/loss due to FW-DRA and BW-DRA together with ISRS and how fast the channel gain/loss due to the FW-DRA and BW-DRA together with ISRS extinguishes along the fibre.

Using the proposed semi-analytical solution of the Raman equations for $\rho(f_i,z)$, the $\text{SNR}_{\text{NLI},i}$ can be obtained as Eq.\eqref{eq:NLI}.
\begin{figure*}[t!]
\vspace*{0.5cm}
\begin{equation} 
\label{eq:NLI}
\end{equation}
\vspace*{-2cm}
\begin{equation*}
	\resizebox{\textwidth}{!}{$%
	\begin{split}
	&\text{SNR}_{\text{NLI},i}^{-1} \approx T^2 \sum_{\substack{0 \leq l_1 + l_2 \leq 1 \\ 0 \leq l_1\textquotesingle + l_2\textquotesingle \leq 1}} \left(\frac{-\tilde{T}_f }{T} \right)^{l_1+l_1\text{\textquotesingle}}\left(\frac{\tilde{T}_b }{T} \right)^{l_2+l_2\text{\textquotesingle}}
	\Bigg( \frac{16}{27}\frac{\pi \gamma^2 P_i^2 n^{1+\epsilon} }{B^2_i\phi_{i} (\alpha_l + \alpha_l\text{\textquotesingle})} \left\{ 2(\kappa_f \kappa_{f}\text{\textquotesingle} + \kappa_b \kappa_b\text{\textquotesingle})  \left[ \mathrm{asinh}{\left(\frac{3 \phi_i B_i^2}{8 \pi \alpha_l}\right)} + \mathrm{asinh}{\left(\frac{3 \phi_i B_i^2}{8 \pi \alpha_l\text{\textquotesingle} }\right)} \right] \right. \\
	&+ 4 \mathrm{ln}\left(\sqrt{\frac{\phi_i L}{2\pi}} B_i  \right) \left[  - (\kappa_f \kappa_b\text{\textquotesingle} +  \kappa_b \kappa_{f}\text{\textquotesingle})  \left( \mathrm{sign}\left(\frac{\alpha_{\text{l}}}{\phi_{i}} \right)  e^{-|\alpha_{\text{l}}L|} +  \mathrm{sign} \left(\frac{\alpha_{\text{l}}\text{\textquotesingle}}{\phi_{i}} \right)  e^{-|\alpha_{\text{l}}\text{\textquotesingle}L|}\right) + (\kappa_f \kappa_b\text{\textquotesingle} - \kappa_b \kappa_{f}\text{\textquotesingle}) \left(  \mathrm{sign}\left( - \phi_i \right)  e^{-|\alpha_{\text{l}}L|} \right. \right.\\
	&+  \left. \left. \left. \mathrm{sign} \left(\phi_i \right)  e^{-|\alpha_{\text{l}}\text{\textquotesingle}L|}\right) \right]  \right\} +\frac{32}{27}\sum_{k=1,k\neq i}^{N_\mathrm{ch}} \frac{n \gamma^2P_k^2}{\phi_{i,k} B_k (\alpha_l + \alpha_l \text{\textquotesingle})}\left\{2(\kappa_f \kappa_{f}\text{\textquotesingle} + \kappa_b \kappa_b\text{\textquotesingle}) \left[\mathrm{atan}\left(\frac{\phi_{i,k}B_i}{2\alpha_l}\right)
	+\mathrm{atan}\left(\frac{\phi_{i,k}B_i}{2\alpha_l\text{\textquotesingle}}\right)\right] \right. \\
	&+ \left. \pi \left[ - (\kappa_f \kappa_b\text{\textquotesingle} +  \kappa_b \kappa_{f}\text{\textquotesingle}) \left(  \mathrm{sign}\left(\frac{\alpha_{\text{l}}}{\phi_{i,k}} \right)  e^{-|\alpha_{\text{l}}L|} +  \mathrm{sign} \left(\frac{\alpha_{\text{l}}\text{\textquotesingle}}{\phi_{i,k}} \right)  e^{-|\alpha_{\text{l}}\text{\textquotesingle}L|}\right)  + (\kappa_f \kappa_b\text{\textquotesingle} - \kappa_b \kappa_{f}\text{\textquotesingle})
	\left(  \mathrm{sign}\left(-\phi_{i,k} \right)  e^{-|\alpha_{\text{l}}L|} +  \mathrm{sign} \left(\phi_{i,k} \right)  e^{-|\alpha_{\text{l}}\text{\textquotesingle}L|}\right) \right] \right\} \Bigg).
	\end{split}$%
	}
\end{equation*}
\vspace*{-0.5cm}
\end{figure*}
In this equation, $T_f = -[P_fC_f(f-\hat{f})]/\alpha_f$, $T_b = - [P_bC_b(f-\hat{f})]/\alpha_b$, $T = 1 + T_f - T_be^{-\alpha_bL}$,  $\alpha_{\text{l}} = \alpha + l_1\alpha_f - l_2\alpha_b$, $\kappa_f = e^{-(\alpha + l_1\alpha_f)L}$, $\kappa_b = e^{-l_2\alpha_b L}$, $\phi_i=-4\pi^2\left(\beta_2+2\pi\beta_3f_i\right)$, $\phi_{i,k}=-4\pi^2\left(f_k-f_i\right)\left[\beta_2+\pi\beta_3\left(f_i+f_k\right)\right]$, $n$ is the number of spans, $P_i$ is the channel launch power with bandwidth $B_i$, $\gamma$ is the nonlinear coefficient, $N_{ch}$ is the number of channels, $\epsilon$ is the coherent factor. The parameters $\beta_2$ and $\beta_3$ are, respectively,
the group velocity dispersion parameter and its linear slope. A detailed derivation of Eq.~\eqref{eq:NLI} can be found in Sec.\ref{derivation}.

\section{Mathematical Derivation of the Closed-form Expression}
\label{derivation}

\subsection{The nonlinear signal-to-noise ratio}

Let $i$ indicate the channel index, the nonlinear signal-to-noise ratio, $\text{SNR}_{\text{NLI},i}$ is given by 
\begin{equation}
\begin{aligned}
   \text{SNR}_{\text{NLI},i} = \frac{P_i}{\eta_n(f_i)P_i^3}, 
\end{aligned}
\label{eq:snr}
\end{equation}
where $P_i$ is the launch power of the channel under test (CUT) and $\eta_n(f_i)$ is the nonlinear coefficient obtained at the end of the $n$-th span. This document is devoted to the calculation of equation \eqref{eq:snr} in closed-form. 

\subsection{The Integral Expression}
\label{sec:integral_expression}

In this section, the integral expressions used to derive the proposed closed-form expressions are presented. $\eta_{GN,n}(f_i)$ in~\eqref{eq:snr}, can be rewritten as
\begin{equation}
\begin{split}
&\eta_n(f_i) \approx \sum_{j=1}^{n} \left[ \frac{P_{i,j}}{P_i} \right]^2 \cdot [\eta_{\text{SPM}_j}(f_i)n^{\epsilon} + \eta_{\text{XPM},j}(f_i)],
\label{eq:XPM_SPM_gauss}
\end{split}
\end{equation}
where $\eta_{\text{SPM}_j}(f_i)$ is the SPM contribution and $\eta_{\text{XPM}_j}(f_i)$ is the total XPM contribution to the NLI both generated in the $j$-th span. $P_{i,j}$ is the power of channel $i$ launched into the $j$-th span, $\epsilon$ is the coherent factor~\cite[Eq.~22]{gnmodel}.
The XPM contribution ($\eta_{\text{XPM},j}(f_i)$) in~\eqref{eq:XPM_SPM_gauss} is obtained by summing over all CUT-interfering pairs present in the transmitted signal, i.e,
\begin{equation}
\begin{split}
&\eta_{\text{XPM},j}(f_i)  = \sum_{k = i, k \neq i}^{N_{ch}} \eta_{\text{XPM}}^{(k)}(f_i),
\label{eq:XPM_gauss}
\end{split}
\end{equation}
where $N_{\text{ch}}$ is the number of WDM channels and $\eta_{\text{XPM}}^{(k)}(f_i)$is the XPM contribution of a single interfering channel k on channel i. 
The XPM and SPM contribution of a single interfering channel are given respectively by~\cite[Eq. 8,9]{closed_gauss_daniel}
\begin{equation}
\begin{split}
&\eta_{\text{XPM}}^{(k)}(f_i) =\frac{32}{27}\frac{\gamma^2}{B_k^2} \left( \frac{P_k}{P_i} \right)^2 \\
& \int_{\frac{-B_i}{2}}^{\frac{B_i}{2}} df_1 \int_{\frac{-B_k}{2}}^{\frac{B_k}{2}} df_2 \ \Pi \left(\frac{f_1 + f_2}{B_k} \right)\left| \mu (f_1+f_i, f_2 + f_k, f_i) \right|^2,
\label{eq:XPM_GN_integral}
\end{split}
\end{equation}
and
\begin{equation}
\begin{split}
\eta_{\text{SPM}}(f_i) = \frac{1}{2}\eta_{\text{XPM}}^{(i)}(f_i),
\label{eq:SPM_GN_integral}
\end{split}
\end{equation}
where $\Pi(x)$ denotes the rectangular function and $B_k$ is the bandwidth of the channel k. $\mu (f_1, f_2, f_i)$ is the so-called link function or FWM efficiency~\cite{gnmodel}, which is given by~\cite[Eq.~4]{isrsgnmodel}   
\begin{equation}
\begin{split}
&\mu\left(f_1,f_2,f_i\right)\\ 
&=  \left| \int_0^L d\zeta \ \sqrt{\frac{\rho(\zeta,f_1) \rho(\zeta,f_2) \rho(\zeta,f_1 + f_2 - f_i)}{\rho(\zeta,f_i)}} e^{j\phi\left(f_1,f_2,f_i\right)\zeta}\right|^2,
\label{eq:link_function_integral}
\end{split}
\end{equation}
where $\phi=-4\pi^2\left(f_1-f_i\right)\left(f_2-f_i\right)\left[\beta_2+\pi\beta_3(f_1+f_2\right)]$, and $\rho(z,f_i)$ is the normalized signal power profile (see Section~\ref{sec:signal_profile}). $\beta_2$ is the group velocity dispersion (GVD) parameter, $\beta_3$ is the
linear slope of the GVD parameter. 

\subsection{The Signal Power Profile}
\label{sec:signal_profile}

The first step in deriving the proposed closed-form expression in this paper is to find an analytical expression which can accurately model the signal power evolution $\rho(z,f_i)$ in the optical fibre. In the presence of distributed Raman amplification (DRA), $\rho(z,f_i)$ is obtained as the solution of the so-called differential Raman equations, given by
\begin{equation}
\begin{aligned}
  &\frac{\partial P_i}{\partial t} =  - \sum_{k=i+1}^{\text{N}_{\text{ch}}} \frac{f_k}{f_i} g(\Delta f) P_k P_i \pm \sum_{p:f_i>f_p} \frac{f_p}{f_i} g(\Delta f) P_p P_i \\
  & + \sum_{k=1}^{i-1} g(\Delta f) P_k P_i \pm \sum_{p:f_i<f_p} g(\Delta f) P_p P_i - \alpha_i P_i,
\end{aligned}
\label{eq:diff_Raman}
\end{equation}
where, $P_i$,$f_i$ is the power and frequency of the CUT, $P_k$, $f_k$ is the power and frequency of the remaining WDM channels, $P_p$, $f_p$ is the power and the frequency of the pumps, $g_r(\Delta f)$ is the polarization averaged, normalized (by the effective core area $A_\text{eff}$) Raman gain spectrum for a frequency separation $\Delta f = |f_i - f_k|$ and $\alpha_i$ is the frequency-dependent attenuation coefficient. Note that the symbol $\pm$ represents the pump under consideration, i.e, $-$ for forward (FW) pump and $+$ for backward (BW) pump configuration.

Eq.\eqref{eq:diff_Raman} does not usually have general analytical solution, unless in some specific cases~\cite{raman_analytical,zirngibl1998analytical,RamanDan}. In order to enable the closed-form expression to be used in any scenario, such that any number of pumps, launch power profiles and bandwidths, a semi-analytical approach is used. This approach consists in using an analytical baseline solution with free parameters. The parameters are then matched using nonlinear least-squares fitting to reproduce the true power profile, which is obtained by numerically solving ~\eqref{eq:diff_Raman}. The following approach was used in~\cite{MLfitting, closed_gauss_daniel, MZraman} to model the ISRS and FW Raman amplification. In this paper for the first time we proposed a semi-analytical solution to account for any scenario, i.e, ISRS, FW and/or BW Raman amplification.

The proposed semi-analytical solution of~\eqref{eq:diff_Raman} is given by
\begin{equation}
\begin{aligned}
&\rho(z,f_i) = e^{-\alpha z} [1 - (C_f P_{f}L_\text{eff} + C_b P_{b} \tilde{L}_\text{eff})(f_i - \hat{f})]
\end{aligned}
\label{eq:Raman_taylor}
\end{equation}
with $L_\text{eff}(\zeta) = (1-e^{-\alpha_f z})/\alpha_f$ and  $\tilde{L}_\text{eff}(\zeta) = (e^{-\alpha_b(L-z)}-e^{-\alpha_b L})/\alpha_b$, where $L$ is the span length, $\alpha$, $\alpha_f$ and $\alpha_b$ are the fibre attenuation at the signal, FW- and BW -DRA wavelengths, respectively, $\hat{f}$ is the average frequency of the FW and BW pumps, $P_{f}$, and $P_{b}$ are the total launch power respectively from the WDM channels together with any FW pumps, and the BW pumps, $C_f$ and $C_b$ is the slope of a linear regression of the normalized Raman gain spectrum. 
The coefficients $\alpha$, $C_f$, $C_b$, $\alpha_f$, $\alpha_b$ are channel-dependent parameters and matched using nonlinear least-squares fitting to reproduce the solution of the Raman differential equations in the presence of DRA. These parameters model respectively the fibre loss, the gain/loss due to FW-DRA and BW-DRA together with ISRS and how fast the channel gain/loss due to the FW-DRA and BW-DRA together with ISRS extinguishes along the fibre.
The derivation of \eqref{eq:Raman_taylor} is given in Appendix~\ref{appA:triangular_approximation}.

\subsection{The Closed-form Expression}
\label{sec:closedform_expression}
This section presents the new close-form expression supporting distributed Raman amplification. The formula is obtained by using the semi-analytical solution of the power evolution as obtained in Section~\ref{sec:signal_profile} to derive a closed-form expression of the NLI as shown in Section \ref{sec:integral_expression}. 

\subsubsection{The link function}

The first step is to derive a closed-form expression of the link function as in~\eqref{eq:link_function_integral}. Let $T_f = \frac{-P_fC_f(f-\hat{f})}{\alpha_f}$, $T_b = \frac{-P_bC_b(f-\hat{f})}{\alpha_b}$, $T = 1 + T_f - T_be^{-\alpha_bL}$, $\alpha_l = \alpha + l_1 \alpha_f - l_2 \alpha_b$, $\kappa_f = e^{-(\alpha + l_1\alpha_f)L}$, $\kappa_b = e^{-l_2\alpha_b L}$. 
The link function is approximate in closed-form as
\begin{equation}
\begin{split}
&\mu\left(f_1,f_2,f_i\right)
\\&\approx \sum_{\substack{0 \leq l_1 + l_2 \leq 1 \\ 0 \leq l_1^\prime + l_2^\prime \leq 1}} \Upsilon \Upsilon^\prime  \left[  \frac{(\kappa_f \kappa_{f}^\prime + \kappa_b \kappa_b^\prime)( \alpha_l \alpha_l^\prime + \phi^2)}{(\alpha_l^2 + \phi^2)(\alpha_l^{\prime 2} + \phi^2)} \right.\\
&
- \frac{(\kappa_f \kappa_b^\prime +  \kappa_b \kappa_{f}^\prime)( \alpha_l \alpha_l^\prime + \phi^2) }{(\alpha_l^2 + \phi^2)(\alpha_l^{\prime 2} + \phi^2)}\cos(\phi L) \\
&\left. + \frac{(\kappa_f \kappa_b^\prime -  \kappa_b \kappa_{f}^\prime)( \alpha_l - \alpha_l^\prime) \phi }{(\alpha_l^2 + \phi^2)(\alpha_l^{\prime 2} + \phi^2)}\sin(\phi L)
\right],
\label{eq:link_function_closed}
\end{split}
\end{equation}
where $\Upsilon$ is given by
\begin{equation}
\begin{split}
&\Upsilon
= T \left(\frac{-\tilde{T}_f }{T} \right)^{l_1} \left(\frac{\tilde{T}_b }{T} \right)^{l_2}.
\label{eq:Upsilon_link_function}
\end{split}
\end{equation}
The proof of \eqref{eq:link_function_closed} is given in Appendix~\ref{appB:link_function}. The coefficient $\Upsilon^\prime$ is respectively the same as the one in~\eqref{eq:Upsilon_link_function} with the indices $l_1$ and $l_2$ replaced by $l_1^\prime$ and $l_2^\prime$. The same is valid for the variables $\alpha_l^\prime$, $\kappa_f^\prime$ and $\kappa_b^\prime$.

\subsubsection{SPM and XPM contributions}
We now focus in obtaining a closed-form expression for the XPM and SPM NLI contributions from GN~model, which are given by~\eqref{eq:XPM_GN_integral} and \eqref{eq:SPM_GN_integral}, respectively. Using~\eqref{eq:link_function_closed} as an analytical solution of the link function, a closed-form expression for the XPM and SPM are given respectively by
\begin{equation}
\begin{split}
&\eta_{\text{XPM}}^{(k)}(f_i) = \sum_{\substack{0 \leq l_1 + l_2 \leq 1 \\ 0 \leq l_1^\prime + l_2^\prime \leq 1}} \Upsilon \Upsilon^\prime \frac{32}{27}\sum_{k=1,k\neq i}^{N_\mathrm{ch}} \frac{n \gamma^2P_k^2}{\phi_{i,k} B_k (\alpha_{\text{l}} + \alpha_{\text{l}}^\prime)}\\
&\cdot\left\{2(\kappa_f \kappa_{f}^\prime + \kappa_b \kappa_b^\prime) \left[\atan\!\left(\frac{\phi_{i,k}B_i}{2\alpha_l}\right)	+\atan\!\left(\frac{\phi_{i,k}B_i}{2\alpha_l^\prime}\right)\right] \right.\\
	&+ \pi \left[ - (\kappa_f \kappa_b^\prime +  \kappa_b \kappa_{f}^\prime) \left(  \sign\!\left(\frac{\alpha_{\text{l}}}{\phi_{i,k}} \right)  e^{-|\alpha_{\text{l}}L|} +  \sign\!\left(\frac{\alpha_{\text{l}}^\prime}{\phi_{i,k}} \right)  e^{-|\alpha_{\text{l}}^\prime L|}\right) \right. \\
 &+ \left. \left. (\kappa_f \kappa_b^\prime - \kappa_b \kappa_{f}^\prime)
	\left(  \sign\!\left(-\phi_{i,k} \right)  e^{-|\alpha_{\text{l}}L|} +  \sign\!\left(\phi_{i,k} \right)  e^{-|\alpha_{\text{l}}^\prime L|}\right) \right] \right\}
\label{eq:XPM_closed}
\end{split}
\end{equation}
and
\begin{equation}
\begin{split}
&\eta_{\text{SPM}}(f_i) = 
 \sum_{\substack{0 \leq l_1 + l_2 \leq 1 \\ 0 \leq l_1^\prime + l_2^\prime \leq 1}} \Upsilon \Upsilon^\prime \frac{16}{27}\frac{\pi \gamma^2 P_i^2 n^{1+\epsilon} }{B^2_i\phi_{i}(\alpha_l + \alpha_l^\prime)} \left\{ 2(\kappa_f \kappa_{f}^\prime + \kappa_b \kappa_b^\prime)  \right. \\
&\cdot \left[\asinh\!{\left(\frac{3 \phi_i B_i^2}{8 \pi\alpha_l}\right)}  + \asinh\!{\left(\frac{3 \phi_i B_i^2}{8 \pi \alpha_l^\prime}\right)} \right] + 4 \ln\!\left(\sqrt{\frac{\phi_i L}{2\pi}} B_i  \right) \\
&\cdot \left[  - (\kappa_f \kappa_b^\prime +  \kappa_b \kappa_{f}^\prime)  \left( \sign\!\left(\frac{\alpha_{\text{l}}}{\phi_{i}} \right)  e^{-|\alpha_{\text{l}}L|} + \sign\!\left(\frac{\alpha_{\text{l}}^\prime}{\phi_{i}} \right)  e^{-|\alpha_{\text{l}}^\prime L|}\right) \right. \\
&+ \left. \left. (\kappa_f \kappa_b^\prime - \kappa_b \kappa_{f}^\prime )\left(  \sign\left( - \phi_i \right)  e^{-|\alpha_{\text{l}}L|}  \sign\left(\phi_i \right)  e^{-|\alpha_{\text{l}}^\prime L|}\right) \right]  \right\},
\label{eq:SPM_closed}
\end{split}
\end{equation}
with $\phi_i=-4\pi^2\left(\beta_2+2\pi\beta_3f_i\right)$ and $\phi_{i,k}=-4\pi^2\left(f_k-f_i\right)\left[\beta_2+\pi\beta_3\left(f_i+f_k\right)\right]$. The proof of \ref{eq:XPM_closed} and \ref{eq:SPM_closed} are given respectively in Appendix~\ref{appC:XPM} and~\ref{appD:SPM}.

\section{Conclusions}

A closed-form expression that can evaluate the NLI in UWB transmission system using distributed Raman amplification was proposed. The approach was enabled by deriving a semi-analytical solution for the channel signal profile evolution along the fibre distance. The formula accounts for both forward and backward amplification, supporting any number of pumps. Using this formula, the NLI is calculated in a few microseconds being suitable for intelligent UWB network planning tools and rapid system and network performance evaluations.

\appendices
\section{Derivation of the Analytical Solution of the normalized signal power profile.}
\label{appA:triangular_approximation}

This section shows the derivation of Eq.~\eqref{eq:Raman_taylor}. We start with Eq.~\eqref{eq:diff_Raman}. The derivation is analogous to \cite{zirngibl1998analytical}. We start by neglecting the energy that is lost whenever a high-frequency photon
is transformed into a low-frequency photon, i.e, $\frac{f_k}{f_i} \approx 1$ and $\frac{f_p}{f_i} \approx 1$. Also, we assume the triangular approximation of the Raman spectrum, i.e, $g_r(\Delta f) \approx C_r \Delta f$, where $C_r$ is the slope of the linear regression (normalized by the effective core area $A_\text{eff}$) and $\Delta f$ is the frequency separation between the channels or the pumps. Under these assumption, \eqref{eq:diff_Raman} can be written as

\begin{equation}
\begin{aligned}
  &\frac{\partial P_i}{\partial z} = \\ &\sum_{k=1}^{N_{ch}} C_r (f_k - f_i) P_k P_i + \sum_{p=1}^{N_p} C_r (f_p - f_i) P_k P_i - \alpha_i P_i = \\
  &  C_r P_i  \left (\sum_{k=1}^{N_{ch}} (f_k - f_i) P_k + \sum_{p=1}^{N_p}(f_p - f_i) P_k \right) - \alpha_i P_i.
\end{aligned}
\label{eq:diff_Raman_app}
\end{equation}
We now write the coupled differential equations into one equation, by replacing the $N_{ch}$ signals and $N_p$ pumps by a signal and pump density spectrum. Also, we replace the
summation by an integration over the entire frequency spectrum
of the signal and the pumps. Thus, \eqref{eq:diff_Raman_app} can be written as

\begin{equation}
\begin{aligned}
  &\frac{d P(z,f)}{d z} = \\
  & C_r P(z,f)  \left (\int_{f_{ch,min}}^{f_{ch,max}} (\Lambda_{ch} - f) P(z,\Lambda_{ch}) \,d\Lambda_{ch} \right.\\
  & \left. + \int_{f_{p,min}}^{f_{p,max}}(\Lambda_p - f) P(z,\Lambda_p) \, d\Lambda_p \right) - \alpha P(z,f).
\end{aligned}
\label{eq:diff_Raman_app_int}
\end{equation}
Now we divide both sides of \eqref{eq:diff_Raman_app_int} per $P(z,f)$ and take the derivative with respect to the frequency $f$,

\begin{equation}
\begin{aligned}
  &\frac{d}{d f} \left ( \frac{d P(z,f)/d z}{P(z,f)} \right )=- C_r \\
  &  \left (\underbrace{\int_{f_{ch,min}}^{f_{ch,max}} P(z,\Lambda_{ch}) \,d\Lambda_{ch}}_\text{$P_{total, ch}$} \right.
  & \left. + \underbrace{\int_{f_{p,min}}^{f_{p,max}} P(z,\Lambda_p) \, d\Lambda_p}_\text{$P_{total, p}$} \right)
\end{aligned}
\label{eq:diff_Raman_app_int2}
\end{equation}
Note that, the integrals represent the total launch power ($P(z)$), i.e, the total launch power of the channels ($P_{total, ch}$) and the pumps ($P_{total, p}$). The total launch power of the channels and the forward pumps ($P_{total, fw}$) must decay with $e^{-\alpha z}$, while for the backward pumps the total launch power ($P_{total, bw}$) decays with $e^{-\alpha(L-z)}$. Thus, \eqref{eq:diff_Raman_app_int2} can be written as

\begin{equation}
\begin{aligned}
  &\frac{d}{d f} \left ( \frac{d P(z,f)/d z}{P(z,f)} \right )= -C_rP(z) =\\
  &- C_r \, (P_{total, ch}e^{-\alpha z} + P_{total, fw}e^{-\alpha z} + P_{total, bw}e^{-\alpha (L-z)})
\end{aligned}
\label{eq:diff_Raman_app_int3}
\end{equation}
Now, in order to apply this equation in more general scenarios, we define separate attenuation for channels and FW pumps ($\alpha_f$) and backward pumps ($\alpha_b$). These parameters model respectively how fast the channel gain/loss due to the FW-DRA and BW-DRA together with ISRS extinguishes along the fibre. We also define separate $C_r$ for each pump configuration, i.e, $C_f$ and $C_b$, respectively for FW and BW pumps. The two parameters models respectively the gain/loss due to FW-DRA and BW-DRA together with ISRS. Finally, by letting $P_f = P_{total, ch} + P_{total, fw}$ and $P_b = P_{total, bw}$, Eq.~\eqref{eq:diff_Raman_app_int3} is rewritten as
\begin{equation}
\begin{aligned}
  &\frac{d}{d f} \left ( \frac{d P(z,f)/d z}{P(z,f)} \right ) =
  &- (C_f P_f e^{-\alpha_f z} + C_bP_be^{-\alpha_b (L-z)})
\end{aligned}
\label{eq:diff_Raman_app_int4}
\end{equation}
Now, we integrate with respect to $z$ and $f$. For the integration in $f$, note that, because of the presence of pumps, the WDM spectra are no longer centred at 0. Without loss of generality, lets consider the center of the spectrum as the average frequencies of the pumps $\hat{f}$. Thus, integrating over $z$ and $f$ yields
\begin{equation}
\begin{aligned}
  &P(z,f)=
  &e^{- [C_f P_{f} L_\text{eff}(f-\hat{f}) + C_b P_b \tilde{L}_\text{eff}(f-\hat{f})]+A(z) + B(f)}
\end{aligned}
\label{eq:diff_Raman_app_int5}
\end{equation}
where $L_\text{eff} = \frac{1-e^{-\alpha_f z}}{\alpha_g}$ and $\tilde{L}_\text{eff} = \frac{e^{-\alpha_b(L-z)}-e^{-\alpha_b L}}{\alpha_b}$, and $A(z)$, $B(f)$ arbitrary functions which their values determined by requiring that $P(z=0,f) = P(0,f)$, which immediately implies that $e^{B(f)}=P(0,f)$. Finally, by requiring $\int P(z,f)\,df = P(z) = P_{total}e^{-\alpha z}$, the value of $e^{A(z)}$ is obtained, leading to equation~\eqref{appB:eq:diff_Raman_app_int6} as
\begin{equation}
\begin{aligned}
&\rho(z,f) = \frac{P(z,f)}{P(0,f)} \\
& = \frac{P_{total}e^{-\alpha z}e^{-(C_fP_{f}L_\text{eff} + C_b P_{b}\tilde{L}_\text{eff})(f-\hat{f})}}{\int G_{\text{Tx}}(\nu)e^{-(C_f P_{f}L_\text{eff} + C_b P_{b}\tilde{L}_\text{eff})\nu} d\nu},
\end{aligned}
\label{appB:eq:diff_Raman_app_int6}
\end{equation}
$G_{Tx}(f)$ is the input signal spectra including the WDM channels and the pumps. Let
$x = C_r(P_{ch,fw}L_\text{eff} + P_{bw}\tilde{L}_\text{eff})$. By assuming that the input power $G_{Tx}(f)$ is uniformly distributed over the optical bandwidth $B$ with power $P_{total}$ we can write, 
\begin{equation}
\begin{aligned}
\int G_{\text{Tx}}(\nu)e^{-x\nu} d\nu = \frac{2P_{total}\sinh{\big(\frac{xB}{2}}\big)}{xB}.
\end{aligned}
\label{appB:eq:diff_Raman_app_int7}
\end{equation}
Replacing \eqref{appB:eq:diff_Raman_app_int7} in \eqref{appB:eq:diff_Raman_app_int6} leads to
\begin{equation}
\begin{aligned}
&\rho(z,f) =e^{-\alpha z} \frac{xB_{ch} e^{-x(f-\hat{f})}}{2\sinh{\big(\frac{xB_{ch}}{2}\big)}}. 
\end{aligned}
\label{appB:eq:diff_Raman_app_int8}
\end{equation}
Finally, by expanding~\eqref{appB:eq:diff_Raman_app_int8} using a 1-st order Taylor approximation around the point $x = 0$, yields
\begin{equation}
\begin{aligned}
&\rho(z,f) = e^{-\alpha z}[1 - x(f - \hat{f})], 
\end{aligned}
\label{appB:eq:diff_Raman_app_int9}
\end{equation}
and Eq.\eqref{eq:Raman_taylor} is obtained concluding the proof.

\section{Derivation of the link function.}
\label{appB:link_function}

Let $x(\zeta) = 1-[C_fP_{f}L_\text{eff}(\zeta) + C_bP_{b}\tilde{L}_\text{eff}(\zeta)](f_i - \hat{f})$ with $L_\text{eff}(\zeta) = \frac{1-e^{-\alpha_f z}}{\alpha_f}$ and  $\tilde{L}_\text{eff}(\zeta) = \frac{e^{-\alpha_b(L-z)}-e^{-\alpha_b L}}{\alpha_b}$. We start by inserting \eqref{eq:Raman_taylor} in \eqref{eq:link_function_integral}, which yields to
\begin{equation}
\begin{split}
&\mu\left(f_1,f_2,f_i\right) =  \left| \int_0^L d\zeta \
e^{-\alpha z} x(\zeta)
e^{j\phi\left(f_1,f_2,f_i\right)\zeta}\right|^2,
\label{appC:link_function_integral1}
\end{split}
\end{equation}
The term $x(\zeta)$ can be written as
\begin{equation}
\begin{split}
&x(\zeta) = 1 - \left[\left(\frac{C_fP_{f}}{\alpha_f}\right) \left(1-e^{-\alpha_f\zeta}\right)  \right.\\
& + \left. \left(\frac{C_bP_{b}}{\alpha_b}\right) e^{-\alpha_b L} \left(e^{\alpha_b \zeta}-1\right)\right](f_i - \hat{f}).
\label{appC:link_function_integral2}
\end{split}
\end{equation}
Let $T_f = \frac{-P_fC_f(f_i-\hat{f})}{\alpha_f}$, $T_b = \frac{-P_bC_b(f_i-\hat{f})}{\alpha_b}$, $T = 1 + T_f - T_be^{-\alpha_bL}$. Thus, the term $x(\zeta)$ is written as 
\begin{equation}
\begin{split}
&x(\zeta) = T[1 - \frac{T_f}{T}e^{-\alpha_f \zeta} + \frac{T_b}{T}e^{-\alpha_b L}e^{\alpha_b \zeta}].
\label{appC:link_function_integral3}
\end{split}
\end{equation}
Eq.\eqref{appC:link_function_integral3} can be conveniently rewritten in terms of a summation using identity \eqref{appE:multinomial_theorem}, which will facilitate all the mathematical derivations,
\begin{equation}
\begin{split}
&x(\zeta) =  T\sum_{\substack{0 \leq l_1 + l_2 \leq 1}} \left( \frac{-T_f}{T} \right)^{l_1} \left( \frac{T_b}{T} \right)^{l_2} e^{-(l_1\alpha_f \zeta + l_2 \alpha_b L - l_2 \alpha_b \zeta)}
\label{appC:link_function_integral4}
\end{split}
\end{equation}
Now, defining 
\begin{equation}
\begin{split}
&\Upsilon
= T \left(\frac{-\tilde{T}_f }{T} \right)^{l_1} \left(\frac{\tilde{T}_b }{T} \right)^{l_2},
\label{appC:link_function_integral5}
\end{split}
\end{equation}
Eq.~\eqref{appC:link_function_integral4} is written as 
\begin{equation}
\begin{split}
&x(\zeta) =  \sum_{\substack{0 \leq l_1 + l_2 \leq 1}} \Upsilon e^{-(l_1\alpha_f \zeta + l_2 \alpha_b L - l_2 \alpha_b \zeta)}.
\label{appC:link_function_integral6}
\end{split}
\end{equation}
Note that $\Upsilon$ is a variable which depends on the indices of the summation. 
Now, inserting \eqref{appC:link_function_integral6} in \eqref{appC:link_function_integral1} 
\begin{equation}
\begin{split}
&\mu\left(f_1,f_2,f_i\right) = \\
& \left|\sum_{\substack{0 \leq l_1 + l_2 \leq 1}} \Upsilon \int_0^L d\zeta \
 e^{-(\alpha \zeta + l_1\alpha_f \zeta + l_2 \alpha_b L - l_2 \alpha_b \zeta) + j \phi \zeta }\right|^2,
\label{appC:link_function_integral7}
\end{split}
\end{equation}
Solving the integral in \eqref{appC:link_function_integral7} yields to 
\begin{equation}
\begin{split}
&\mu\left(f_1,f_2,f_i\right) =  \left|\sum_{\substack{0 \leq l_1 + l_2 \leq 1}} \Upsilon \frac{e^{-(\alpha + l_1\alpha_f)L + j \phi L} - e^{-l_2\alpha_b L}}{-(\alpha + l_1 \alpha_f - l_2 \alpha_b) + j \phi}\right|^2.
\label{appC:link_function_integral8}
\end{split}
\end{equation}
Now, let define $\alpha_l = \alpha + l_1 \alpha_f - l_2 \alpha_b$, $\kappa_f = e^{-(\alpha + l_1\alpha_f)L}$ and $\kappa_b = e^{-l_2\alpha_b L}$. Eq.~\eqref{appC:link_function_integral8} can then be written as 
\begin{equation}
\begin{split}
&\mu\left(f_1,f_2,f_i\right) =  \left|\sum_{\substack{0 \leq l_1 + l_2 \leq 1}} \Upsilon \frac{\kappa_f e^{j \phi L} - \kappa_b}{- \alpha_l + j \phi}\right|^2.
\label{appC:link_function_integral9}
\end{split}
\end{equation}
The last step of the derivation is to calculate the modulus of \eqref{appC:link_function_integral9}. Using the identity \eqref{appE:modulo_complex_number} we can write \eqref{appC:link_function_integral9} as
\begin{equation}
\begin{split}
&\mu\left(f_1,f_2,f_i\right)\\ 
&=  \left ( \sum_{\substack{0 \leq l_1 + l_2 \leq 1}} \Upsilon \frac{\kappa_f e^{j \phi L} - \kappa_b}{- \alpha_l + j \phi} \right) 
\left ( \sum_{\substack{0 \leq l_1^\prime + l_2^\prime \leq 1}} \Upsilon^\prime \frac{\kappa_f^\prime e^{j \phi L} - \kappa_b^\prime}{- \alpha_l^\prime - j \phi}\right).
\label{appC:link_function_integral10}
\end{split}
\end{equation}
Finally, performing the multiplication in Eq.\eqref{appC:link_function_integral10} together with the identity \eqref{appE:sum_modulo_complex_number} yields to Eq.~\eqref{eq:link_function_closed}, concluding the proof.

\section{Derivation of the XPM contribution.}
\label{appC:XPM}

This section shows the derivation of \eqref{eq:XPM_closed}. We start by approximating the phase mismatch term in~\eqref{eq:link_function_integral}.  For the XPM contribution, let $\Delta f = f_k - f_i$ be the frequency separation between channels $k$ and $i$. Assuming that frequency
separation is much larger than half of the bandwidth of channel $k$ ($|\Delta f| \gg \frac{B_k}{2}$), we can make the assumption that $f_2 + \Delta f \approx \Delta f$. Also, we assume that the dispersion slope $\beta_3$ is constant over the channel bandwidth. Thus, the phase mismatch term can be approximated as~\cite[Eq. 15]{closed_gauss_daniel},

\begin{equation}
\begin{split}
&\phi(f_1+f_i,f_2+f_k,f_i)=\\
& = -4\pi^2 f_1 \Delta f \left[\beta_2+\pi\beta_3(f_1+f_2+f_i+f_k\right)]\\
&\approx-4\pi^2 (f_k-f_i) \left[\beta_2+\pi\beta_3(f_i+f_k\right)]f_1\\
&=\phi_{i,k}f_1,
\label{appD:mismatch_term}
\end{split}
\end{equation}
with $\phi_{i,k} = -4\pi (f_k-f_i) \left[\beta_2+\pi\beta_3(f_i+f_k\right)]$. The most impacted channels by this approximation is the ones near the CUT. The error relative to this approximation is given by \cite[Eq. 25]{closed_gauss_daniel}.
 
Now, we consider Eq.\eqref{eq:XPM_GN_integral}. For notation brevity, we will omit the factor $\frac{32}{27}\frac{\gamma^2}{B_k^2} \left( \frac{P_k}{P_i} \right)^2$. Also, the term $\Pi \left(\frac{f_1 + f_2}{B_k} \right)$ is neglected - this is equivalent of approximating the integration domain of the GN model to a rectangle~\cite{gnmodel}. Because of the approximation in Eq.\eqref{appD:mismatch_term}, $\phi$ no longer depends on $f_2$, and the double integral in~\eqref{eq:XPM_GN_integral} turns to be a single integral.  Thus, inserting Eq.\eqref{eq:link_function_closed} in Eq.\eqref{eq:XPM_GN_integral}, we can identify, three terms as follows

\begin{equation}
\begin{split}
&\eta_{\text{XPM}}^{(k)}(f_i) = \sum_{\substack{0 \leq l_1 + l_2 \leq 1 \\ 0 \leq l_1^\prime + l_2^\prime \leq 1}} \Upsilon \Upsilon^\prime  [(\kappa_f \kappa_{f}^\prime + \kappa_b \kappa_b^\prime)\eta_{\text{XPM,main}}^{(k)}(f_i) 
\\&- (\kappa_f \kappa_b^\prime +  \kappa_b \kappa_{f}^\prime)\eta_{\text{XPM,cos}}^{(k)}(f_i) + (\kappa_f \kappa_b^\prime -  \kappa_b \kappa_{f}^\prime)\eta_{\text{XPM,sin}}^{(k)}(f_i)].
\label{appD:XPM_GN_integral1}
\end{split}
\end{equation}
with
\begin{equation}
\begin{split}
&\eta_{\text{XPM,main}}^{(k)}(f_i) = 2B_k\int_{0}^{\frac{B_i}{2}} df_1   \frac{\alpha_l \alpha_l ^\prime + \phi_{i,k}^2f_1^2}{(\alpha_l^2 + \phi_{i,k}^2f_1^2)(\alpha_l^{\prime 2} + \phi_{i,k}^2f_1^2)},
\label{appD:XPM_GN_integral2}
\end{split}
\end{equation}
\begin{equation}
\begin{split}
&\eta_{\text{XPM,cos}}^{(k)}(f_i) \\&= 2B_k\int_{0}^{\frac{B_i}{2}} df_1  \frac{\alpha_l \alpha_l ^\prime + \phi_{i,k}^2f_1^2}{(\alpha_l^2 + \phi_{i,k}^2f_1^2)(\alpha_l^{\prime 2} + \phi_{i,k}^2f_1^2)}\cos(\phi_{i,k} L)
\label{appD:XPM_GN_integral3}
\end{split}
\end{equation}
and
\begin{equation}
\begin{split}
&\eta_{\text{XPM,sin}}^{(k)}(f_i) \\&= 2B_k\int_{0}^{\frac{B_i}{2}} df_1  \frac{(\alpha_l - \alpha_l) ^\prime \phi_{i,k}f_1 }{(\alpha_l^2 + \phi_{i,k}^2f_1^2)(\alpha_l^{\prime 2} + \phi_{i,k}^2f_1^2)}\sin(\phi_{i,k} L).
\label{appD:XPM_GN_integral4}
\end{split}
\end{equation}
In the following, the above three integrals are solved. Eq.\eqref{appD:XPM_GN_integral2} is solving using identity \eqref{appE:integral1} as
\begin{equation}
\begin{split}
&\eta_{\text{XPM,main}}^{(k)}(f_i) =\frac{2B_k}{\phi_{i,k}(\alpha_l + \alpha_l^\prime)}\\
&\left[\arctan\left(\frac{\phi_{i,k}B_i}{2\alpha_l}\right)
+ \arctan\left(\frac{\phi_{i,k}B_i}{2\alpha_l^\prime}\right)\right],
\label{appD:XPM_GN_integral5}
\end{split}
\end{equation}
Eqs.\eqref{appD:XPM_GN_integral3} and \eqref{appD:XPM_GN_integral4} do not have analytical solutions in its current form. In order to derive an analytical solution, we extend the channel bandwidth $B_i \rightarrow \infty$ and solve it using identities~\eqref{appE:integral4} and~\eqref{appE:integral5}, yielding to
\begin{equation}
\begin{split}
&\eta_{\text{XPM,cos}}^{(k)}(f_i) =\frac{\pi B_k}{\phi_{i,k}(\alpha_l + \alpha_l^\prime)}\\
&\cdot \left[ e^{-|\alpha_l L|}\sign\left(\frac{\phi_{i,k}}{\alpha_l}\right) + e^{-|\alpha_l^\prime L|}\sign\left(\frac{\phi_{i,k}}{\alpha_l^\prime}\right) \right]
\label{appD:XPM_GN_integral6}
\end{split}
\end{equation}
and
\begin{equation}
\begin{split}
&\eta_{\text{XPM,sin}}^{(k)}(f_i) =\frac{\pi B_k}{\phi_{i,k}(\alpha_l + \alpha_l^\prime)}\\
&\cdot \left[ e^{-|\alpha_l L|}\sign\left(-\phi_{i,k}\right) + e^{-|\alpha_l^\prime L|}\sign\left(\phi_{i,k}\right) \right]
\label{appD:XPM_GN_integral7}
\end{split}
\end{equation}
Finally, by inserting Eqs. \eqref{appD:XPM_GN_integral5}, \eqref{appD:XPM_GN_integral6} and \eqref{appD:XPM_GN_integral7} in \eqref{appD:XPM_GN_integral1} together with the pre-factor $\frac{32}{27}\frac{\gamma^2}{B_k^2} \left( \frac{P_k}{P_i} \right)^2$, Eq.\eqref{eq:XPM_closed} is obtained concluding the proof.

\section{Derivation of the SPM contribution.}
\label{appD:SPM}

This section shows the derivation of \eqref{eq:SPM_closed}. We start by approximating the phase mismatch term. We assume that the dispersion slope $\beta_3$ is constant over the channel bandwidth. Thus, the phase mismatch term can be approximated as  
\begin{equation}
\begin{split}
&\phi(f_1+f_i,f_2+f_i,f_i)=\\
&-4\pi^2f_1f_2\left[\beta_2+\pi\beta_3(f_1+f_2-2f_i\right)]\\
&\approx-4\pi^2f_1f_2(\beta_2+2\pi\beta_3f_i)\\
&=\phi_{i}f_1f_2,
\label{appD:mismatch_term_SPM}
\end{split}
\end{equation}
with $\phi_{i} = -4\pi^2(\beta_2+2\pi\beta_3f_i)$. 

Now, using \eqref{eq:SPM_GN_integral} together with~\eqref{eq:XPM_GN_integral} and \eqref{eq:link_function_closed} with $k=i$, and omitting the pre-factor of  $\frac{16}{27}\frac{\gamma^2}{B_i^2}$, we can write
\begin{equation}
\begin{split}
&\eta_{\text{SPM}}(f_i) = \sum_{\substack{0 \leq l_1 + l_2 \leq 1 \\ 0 \leq l_1^\prime + l_2^\prime \leq 1}} \Upsilon \Upsilon^\prime  [(\kappa_f \kappa_{f}^\prime + \kappa_b \kappa_b^\prime)\eta_{\text{SPM,main}}(f_i) 
\\&- (\kappa_f \kappa_b^\prime +  \kappa_b \kappa_{f}^\prime)\eta_{\text{SPM,cos}}(f_i) + (\kappa_f \kappa_b^\prime -  \kappa_b \kappa_{f}^\prime)\eta_{\text{SPM,sin}}(f_i)],
\label{appD:SPM_GN_integral1}
\end{split}
\end{equation}
where $\eta_{\text{SPM}}(f_i)$, $\eta_{\text{SPM,cos}}(f_i)$ and $\eta_{\text{SPM,sin}}(f_i)$ are given respectively by

\begin{equation}
\begin{split}
&\eta_{\text{SPM,main}}(f_i)\\& = \int_{-\frac{B_i}{2}}^{\frac{B_i}{2}} df_1
\int_{-\frac{B_i}{2}}^{\frac{B_i}{2}} df_2
\frac{\alpha_l \alpha_l ^\prime + \phi_i^2f_1^2f_2^2}{(\alpha_l^2 + \phi_i^2f_1^2f_2^2)(\alpha_l^{\prime 2} + \phi_i^2f_1^2f_2^2)},
\label{appD:SPM_GN_integral2}
\end{split}
\end{equation}
\begin{equation}
\begin{split}
&\eta_{\text{SPM,cos}}(f_i) \\&= \int_{-\frac{B_i}{2}}^{\frac{B_i}{2}} df_1  \int_{-\frac{B_i}{2}}^{\frac{B_i}{2}} df_2\frac{\alpha_l \alpha_l ^\prime + \phi_i^2f_1^2f_2^2}{(\alpha_l^2 + \phi_i^2f_1^2f_2^2)(\alpha_l^{\prime 2} + \phi_i^2f_1^2f_2^2)}\cos(\phi L)
\label{appD:SPM_GN_integral3}
\end{split}
\end{equation}
and
\begin{equation}
\begin{split}
&\eta_{\text{SPM,sin}}(f_i) \\&= \int_{-\frac{B_i}{2}}^{\frac{B_i}{2}} df_1 \int_{-\frac{B_i}{2}}^{\frac{B_i}{2}} df_2 \frac{(\alpha_l - \alpha_l) ^\prime \phi_if_1f_2 }{(\alpha_l^2 + \phi_i^2f_1^2f_2^2)(\alpha_l^{\prime 2} + \phi_i^2f_1^2f_2^2)}\sin(\phi L).
\label{appD:SPM_GN_integral4}
\end{split}
\end{equation}
Note that, similar to Appendix~\ref{appC:XPM}, the term $\Pi \left(\frac{f_1 + f_2}{B_i} \right)$ is neglected. In the following the three integrals above are solved. The integral in~\eqref{appD:SPM_GN_integral2} is rewritten in polar coordinates $(r,\varphi)$ as
\begin{equation}
\begin{split}
&\eta_{\text{SPM,main}}(f_i) \approx 4\int_{0}^{\sqrt{\frac{3}{\pi}}\frac{B_i}{2}} dr \int_{0}^{\frac{\pi}{2}} d\varphi  \ \\
&\cdot \frac{r\left[\tilde{\alpha} \tilde{\alpha} ^\prime + \frac{\phi_i^2}{4} (r^4\sin^2{(\varphi)}) \right]}{\left[\tilde{\alpha}^2 + \frac{\phi_i^2}{4} (r^4\sin^2{(\varphi)}) \right]\left[\tilde{\alpha} ^{\prime 2} + \frac{\phi_i^2}{4} (r^4\sin^2{(\varphi)}) \right]},
\label{appD:SPM_GN_integral5}
\end{split}
\end{equation}
where it was used the relations $f_1 = r\cos{(\varphi /2)}$, $f_2 = r\sin{(\varphi / 2)}$ and $ \sin{(\varphi / 2)}\cos{(\varphi / 2)} = \frac{\sin{(\varphi)}}{2}$. Also the integration domain of \eqref{eq:SPM_GN_integral} was approximated by a circular domain such that the area of both domains are equal \cite[Fig.3]{closed_gauss_daniel}. This yields the variation of the radius in the outer integral as shown in \eqref{appD:SPM_GN_integral5}. 
The inner integral in \eqref{appD:SPM_GN_integral5} can be solved using identity \eqref{appE:integral2}, yielding to
\begin{equation}
\begin{split}
&\eta_{\text{SPM,main}}(f_i) \approx 4\int_{0}^{\sqrt{\frac{3}{\pi}}\frac{B_i}{2}} dr\\
&\cdot  \frac{ r \pi}{\alpha_l +\alpha_l^\prime }\left[\frac{1}{\sqrt{4 \alpha_l^2 + \phi_i^2 r^4}} + \frac{1}{\sqrt{4 \alpha_l^{\prime 2} + \phi_i^2 r^4}}\right],
\label{appD:SPM_GN_integral6}
\end{split}
\end{equation}
This integral can be rewritten as:
\begin{equation}
\begin{split}
&\eta_{\text{SPM,main}}(f_i) =  \frac{ 2\pi}{\alpha_l +\alpha_l^\prime }\int_{0}^{\sqrt{\frac{3}{\pi}}\frac{B_i}{2}} dr\\
&\cdot \left[\frac{r}{ \alpha_l\sqrt{1 + \frac{\phi_i^2 r^4}{4 \alpha_l^2}}} + \frac{r}{ \alpha_l^\prime\sqrt{1 + \frac{\phi_i^2 r^4}{4 \alpha_l^{\prime 2}}}}\right],
\label{appD:SPM_GN_integral7}
\end{split}
\end{equation}
The integral in \eqref{appD:SPM_GN_integral7} is solved using identity \eqref{appE:integral3} as,
\begin{equation}
\begin{split}
&\eta_{\text{SPM,main}}(f_i) =  \frac{ 2\pi}{\phi_i(\alpha_l +\alpha_l^\prime)} \\
&\cdot
\left[\asinh{\left(\frac{3 \phi_i B_i^2}{8 \pi \alpha_l } \right)} + \asinh{\left(\frac{3 \phi_i B_i^2}{8 \pi \alpha_l^\prime } \right)} \right].
\label{appD:SPM_GN_integral8}
\end{split}
\end{equation}

To solve the integrals in Eqs.\eqref{appD:SPM_GN_integral3} and \eqref{appD:SPM_GN_integral4}, a similar approach use in~\cite{Semrau:17} are used. The integrals are converted in hyperbolic coordinates using the relations $\nu_1 = \sqrt{f_1 f_2}$, $\nu_2 = -\frac{1}{2} \ln \left(\frac{f_1}{f_2}\right)$, $f_1 = \nu_1 e^{\nu_2}$ and $f_2 = \nu_1 e^{-\nu_2}$~\cite[Sec.VIII-A]{gnmodel}; this change of coordinates yields a one-dimensional integral in $\nu_1$. We also use the change of variable $\nu^2 = \nu_1$~\cite{Semrau:17} to rewrite Eqs.\eqref{appD:SPM_GN_integral3} and \eqref{appD:SPM_GN_integral4} as
\begin{equation}
\begin{split}
&\eta_{\text{SPM,cos}}(f_i) = \\&8\int_{0}^{\frac{B_i}{2}} d\nu \ln\left(\frac{B_i}{2\sqrt{\nu}}\right) \frac{\alpha_l \alpha_l ^\prime + \phi_i^2\nu^2}{(\alpha_l^2 + \phi_i^2\nu^2)(\alpha_l^{\prime 2} + \phi_i^2\nu^2)}\cos(\phi_i L)
\label{appD:SPM_GN_integral9}
\end{split}
\end{equation}
and
\begin{equation}
\begin{split}
&\eta_{\text{SPM,sin}}(f_i) =\\ &8\int_{0}^{\frac{B_i}{2}} d\nu \ln\left(\frac{B_i}{2\sqrt{\nu}}\right) \frac{(\alpha_l - \alpha_l ^\prime) \phi_i\nu}{(\alpha_l^2 + \phi_i^2\nu^2)(\alpha_l^{\prime 2} + \phi_i^2\nu^2)}\sin(\phi_i L),
\label{appD:SPM_GN_integral10}
\end{split}
\end{equation}
The integrals in Eqs.\eqref{appD:SPM_GN_integral9} and \eqref{appD:SPM_GN_integral10} do not have analytical solution in its current form. In order to obtain an integral that yields an analytical solution we evaluate the logarithm functions in the point $\nu = \frac{\pi}{2\phi_i L}$, where this point was chosen such that the cosine function achieves its minima and the sine function achieves its maxima. This yields to  
\begin{equation}
\begin{split}
&\eta_{\text{SPM,cos}}(f_i) = \\&8\ln\left(\sqrt{\frac{\phi_i L}{2 \pi}}B_i\right)\int_{0}^{\frac{B_i}{2}} d\nu \frac{\alpha_l \alpha_l ^\prime + \phi_i^2\nu^2}{(\alpha_l^2 + \phi_i^2\nu^2)(\alpha_l^{\prime 2} + \phi_i^2\nu^2)}\cos(\phi_i L)
\label{appD:SPM_GN_integral11}
\end{split}
\end{equation}
and
\begin{equation}
\begin{split}
&\eta_{\text{SPM,sin}}(f_i) =\\ &8\ln\left(\sqrt{\frac{\phi_i L}{2 \pi}}B_i\right)\int_{0}^{\frac{B_i}{2}} d\nu  \frac{(\alpha_l - \alpha_l ^\prime) \phi_i\nu}{(\alpha_l^2 + \phi_i^2\nu^2)(\alpha_l^{\prime 2} + \phi_i^2\nu^2)}\sin(\phi_i L).
\label{appD:SPM_GN_integral12}
\end{split}
\end{equation}
The integrals in Eqs.\eqref{appD:SPM_GN_integral11} and \eqref{appD:SPM_GN_integral12} can now be solved similar to Appendix~\ref{appC:XPM}, i.e, by letting $B_i \rightarrow \infty$. This yields to 
\begin{equation}
\begin{split}
&\eta_{\text{SPM,cos}}(f_i) = \\&4\pi\ln\left(\sqrt{\frac{\phi_i L}{2 \pi}}B_i\right)\left[ e^{-|\alpha_l L|}\sign\left(\frac{\phi_i}{\alpha_l}\right) + e^{-|\alpha_l^\prime L|}\sign\left(\frac{\phi_i}{\alpha_l^\prime}\right) \right]
\label{appD:SPM_GN_integral13}
\end{split}
\end{equation}
and
\begin{equation}
\begin{split}
&\eta_{\text{SPM,sin}}(f_i) =\\ &4\pi\ln\left(\sqrt{\frac{\phi_i L}{2 \pi}}B_i\right) \left[ e^{-|\alpha_l L|}\sign\left(-\phi_i\right) + e^{-|\alpha_l^\prime L|}\sign\left(\phi_i\right) \right].
\label{appD:SPM_GN_integral14}
\end{split}
\end{equation}
Finally, by inserting \eqref{appD:SPM_GN_integral8}, \eqref{appD:SPM_GN_integral13} and \eqref{appD:SPM_GN_integral14} in \eqref{appD:SPM_GN_integral1} together with the pre-factor of  $\frac{16}{27}\frac{\gamma^2}{B_i^2}$, Eq.\eqref{eq:SPM_closed} is obtained concluding the proof.

\section{Mathematical Identities}
\label{appE:Mathematical_identities}


\begin{multline}
(x + y + z)^i = \\ \sum_{0 \leq 1_1 + 1_2 \leq i} \frac{i!}{l_1! l_2!(i-l_1-1_2)!} x^{l_1}y^{l_2}z^{i-l_1-l_2}.
\label{appE:multinomial_theorem}
\end{multline}

\begin{equation}
\begin{split}
|z_k|^2 = \Re{( z_k \cdot \overline{z}_k)} = z_k \cdot \overline{z}_k. 
\label{appE:modulo_complex_number}
\end{split}
\end{equation}

\begin{equation}
\begin{split}
z_i \cdot \overline{z}_j +  z_j \cdot \overline{z}_i  = 2\Re{( z_i \cdot \overline{z}_j)}, \text{    } j<i.
\label{appE:sum_modulo_complex_number}
\end{split}
\end{equation}

\begin{equation}
\begin{split}
&\int_{0}^{X}dx \ \frac{ab + c^2x^2}{(a^2 + c^2x^2)(b^2 + c^2x^2)} \\
&=\frac{1}{c(a+b)}\left[\arctan\left(\frac{cx}{a}\right) + \arctan\left(\frac{cx}{b}\right)\right].
\label{appE:integral1}
\end{split}
\end{equation}

\begin{equation}
\begin{split}
&\int_{0}^{\frac{\pi}{2}}dx \ \frac{ab + c^2\sin^2{(x)}}{[a^2 + c^2\sin^2{(x)}][b^2 + c^2\sin^2{(x)}]} \\
&= \frac{\pi}{2(a+b)}\left(\frac{1}{\sqrt{a^2 + c^2}} + \frac{1}{\sqrt{b^2 +c^2 }}\right).
\label{appE:integral2}
\end{split}
\end{equation}

\begin{equation}
\begin{split}
&\int_{0}^{X}dx \ \frac{x}{\sqrt{1 + d^2x^4}}
= \frac{1}{2d}\asinh{(dX^2)}.
\label{appE:integral3}
\end{split}
\end{equation}

\begin{equation}
\begin{split}
&\int_{0}^{\infty}dx \ \frac{ab + c^2x^2}{(a^2 + c^2x^2)  (b^2 + c^2x^2)} \cos(cxL) \\
&= \frac{\pi}{2} \frac{e^{-|aL|}\sign(c/a) + e^{-|bL|}\sign(c/b)}{c(a + b)}.
\label{appE:integral4}
\end{split}
\end{equation}

\begin{equation}
\begin{split}
&\int_{0}^{\infty}dx \ \frac{(a-b)cx}{(a^2 + c^2x^2)  (b^2 + c^2x^2)}\sin(cxL)  \\
&= \frac{\pi}{2} \frac{e^{-|aL|}\sign(-c) + e^{-|bL|}\sign(c)}{c(a + b)}.
\label{appE:integral5}
\end{split}
\end{equation}

\bibliographystyle{IEEEbib}
\bibliography{main}

\end{document}